\documentclass{article}
\usepackage{LaThuileFPSpro}
\def\NIMA#1#2#3{Nucl. Inst. Methods {\bf A#1} (#2) #3}
\begin{document}
\title{
  New results from the NA48/2 experiment at CERN SPS:
  radiative nonleptonic kaon decays
  }
\author{
  Evgueni Goudzovski\\
  {\em INFN sezione di Pisa, Largo B. Pontecorvo 3, Pisa, 56127 Italy}
  }
\maketitle

\baselineskip=11.6pt

\begin{abstract}
The NA48/2 experiment at the CERN SPS carried out data taking in
2003 and 2004. Analysis of the selected data samples of 7,146
$K^\pm\to\pi^\pm e^+e^-$ decay candidates with 0.6\% background,
1,164 $K^\pm\to\pi^\pm\gamma\gamma$ candidates with 3.3\%
background, and 120 $K^\pm\to\pi^\pm\gamma e^+e^-$ candidates with
6.1\% background allowed precise measurements of branching fractions
and other characteristics of these rare kaon decays.
\end{abstract}
\newpage
\section*{Introduction}
Radiative nonleptonic kaon decays represent a source of information
on the structure of the weak interactions at low energies, and
provide crucial tests of the Chiral Perturbation Theory (ChPT). The
current paper presents new results related to study of the
$K^\pm\to\pi^\pm e^+e^-$, $K^\pm\to\pi^\pm\gamma\gamma$, and
$K^\pm\to\pi^\pm\gamma e^+e^-$ decays by the NA48/2 experiment at
the CERN SPS.

The flavour-changing neutral current process $K^\pm\to\pi^\pm
e^+e^-$, induced at one-loop level in the Standard Model and highly
suppressed by the GIM mechanism, has been described by the
ChPT\cite{ek87}; several models predicting the form factor
characterizing the dilepton invariant mass spectrum and the decay
rate have been proposed\cite{da98,du06}. The decay is fairly well
explored experimentally: it was first studied at CERN\cite{bl75},
followed by BNL E777\cite{al92} and E865\cite{ap99} measurements.

The $K^\pm\to\pi^\pm\gamma\gamma$ and $K^\pm\to\pi^\pm\gamma e^+e^-$
decays similarly arise at one-loop level in the ChPT. The decay
rates and spectra have been computed at leading and next-to-leading
orders\cite{da96,ga99}, and strongly depend on a single
theoretically unknown parameter $\hat c$. The experimental knowledge
of these processes is rather poor: before the NA48/2 experiment,
only a single observation of 31 $K^\pm\to\pi^\pm\gamma\gamma$
candidates was made\cite{ki97}, while the $K^\pm\to\pi^\pm\gamma
e^+e^-$ decay was not observed at all.

The paper is organized as follows. In Section 1, a description of
the NA48/2 experiment is given. Section 2 is devoted to a rather
detailed description of the $K^\pm\to\pi^\pm e^+e^-$ analysis and
its preliminary results, which is the main topic of the paper.
Section 3 briefly presents the preliminary results of the
$K^\pm\to\pi^\pm\gamma\gamma$ analysis; a more detailed discussion
is reserved for the Moriond QCD 2008 conference. Section 4 briefly
presents the final results of the $K^\pm\to\pi^\pm\gamma e^+e^-$
analysis, which have recently been published\cite{ba08}. Finally the
conclusions follow.

\section{The NA48/2 experiment}

\begin{figure}[t]
  \includegraphics{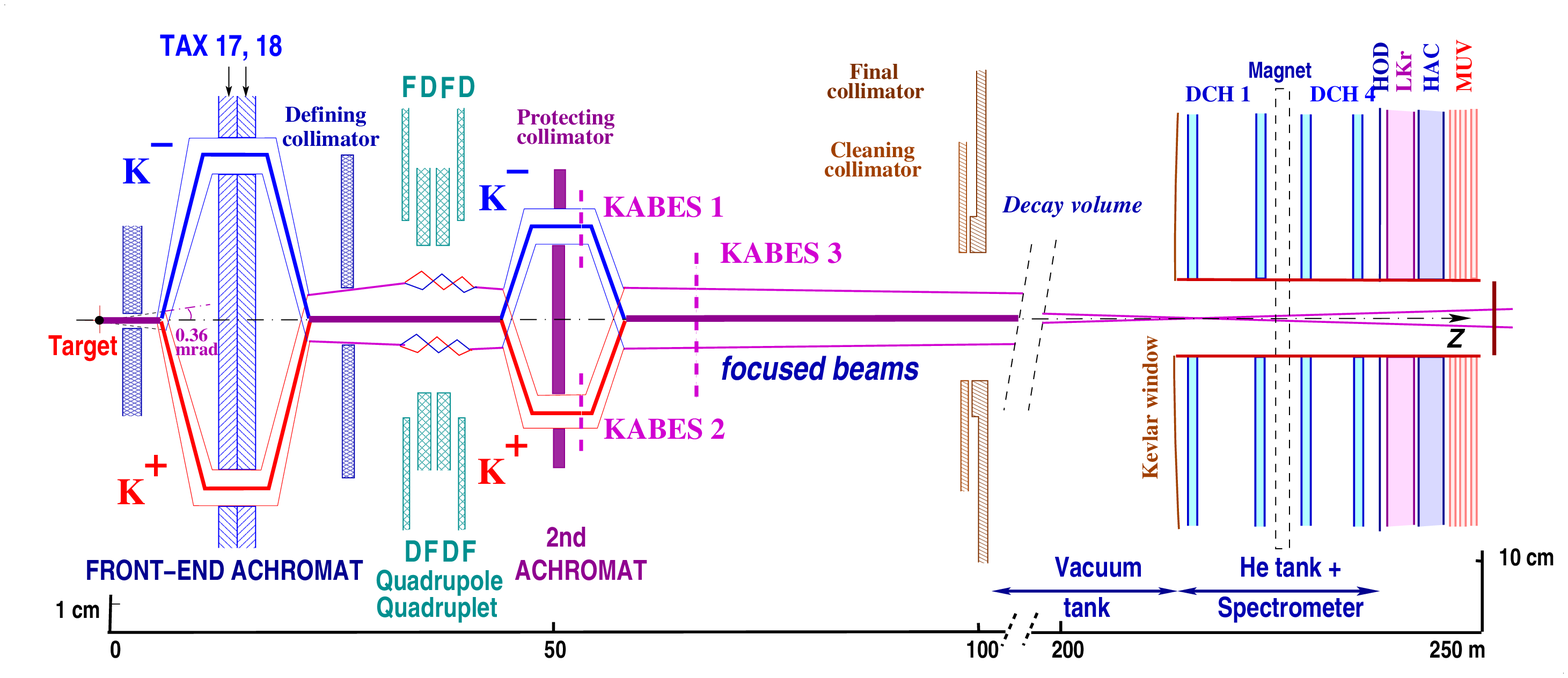}
    \vspace{50mm}
  \caption{\it
Schematic lateral view of the NA48/2 beam line (TAX17,18: motorized
beam dump/collimators used to select the momentum of the $K^+$ and
$K^-$ beams; FDFD/DFDF: focusing set of quadrupoles, KABES1--3: kaon
beam spectrometer stations), decay volume and detector (DCH1--4:
drift chambers, HOD: hodoscope, LKr: EM calorimeter, HAC: hadron
calorimeter, MUV: muon veto). The vertical scales are different in
the two parts of the figure.
    \label{fig:beams} }
\end{figure}

The NA48/2 experiment, designed to excel in charge asymmetry
measurements\cite{ba07}, is based on simultaneous $K^+$ and $K^-$
beams produced by 400 GeV/$c$ primary SPS protons impinging at zero
incidence angle on a beryllium target of 40 cm length and 2 mm
diameter. Charged particles with momentum $(60\pm3)$ GeV/$c$ are
selected by an achromatic system of four dipole magnets with zero
total deflection (`achromat'), which splits the two beams in the
vertical plane and then recombines them on a common axis. Then the
beams pass through a defining collimator and a series of four
quadrupoles designed to produce focusing of the beams towards the
detector. Finally the two beams are again split in the vertical
plane and recombined in a second achromat. The layout of the beams
and detectors is shown schematically in Fig.~\ref{fig:beams}.

The beams then enter the decay volume housed in a 114 m long
cylindrical vacuum tank with a diameter of 1.92 m for the first 65
m, and 2.4 m for the rest. Both beams follow the same path in the
decay volume: their axes coincide within 1~mm, while the transverse
size of the beams is about 1~cm. With $7\times 10^{11}$ protons
incident on the target per SPS spill of $4.8$~s duration, the
positive (negative) beam flux at the entrance of the decay volume is
$3.8\times 10^7$ ($2.6\times 10^7$) particles per pulse, of which
$5.7\%$ ($4.9\%$) are $K^+$ ($K^-$). The $K^+/K^-$ flux ratio is
about $1.8$. The fraction of beam kaons decaying in the decay volume
at nominal momentum is $22\%$.

The decay volume is followed by a magnetic spectrometer housed in a
tank filled with helium at nearly atmospheric pressure, separated
from the vacuum tank by a thin ($0.31\%X_0$) Kevlar composite
window. A thin-walled aluminium beam pipe of 16~cm outer diameter
traversing the centre of the spectrometer (and all the following
detectors) allows the undecayed beam particles and the muon halo
from decays of beam pions to continue their path in vacuum. The
spectrometer consists of four drift chambers (DCH): DCH1, DCH2
located upstream, and DCH3, DCH4 downstream of a dipole magnet. The
magnet provides a horizontal transverse momentum kick $\Delta
p=120~{\rm MeV}/c$ for charged particles. The DCHs have the shape of
a regular octagon with a transverse size of about 2.8 m and a
fiducial area of about 4.5 m$^2$. Each chamber is composed of eight
planes of sense wires arranged in four pairs of staggered planes
oriented horizontally, vertically, and along each of the two
orthogonal $45^\circ$ directions. The spatial resolution of each DCH
is $\sigma_x=\sigma_y=90~\mu$m. The nominal spectrometer momentum
resolution is $\sigma_p/p = (1.02 \oplus 0.044\cdot p)\%$ ($p$ in
GeV/$c$).

The magnetic spectrometer is followed by a plastic scintillator
hodoscope (HOD) used to produce fast trigger signals and to provide
precise time measurements of charged particles. The hodoscope has a
regular octagonal shape with a transverse size of about 2.4~m. It
consists of a plane of horizontal and a plane of vertical
strip-shaped counters. Each plane consists of 64 counters arranged
in four quadrants. Counter widths (lengths) vary from 6.5 cm (121
cm) for central counters to 9.9 cm (60 cm) for peripheral ones.

The HOD is followed by a liquid krypton electromagnetic calorimeter
(LKr)\cite{ba96} used for photon detection and particle
identification. It is an almost homogeneous ionization chamber with
an active volume of 7 m$^3$ of liquid krypton, segmented
transversally into 13248 projective cells, 2$\times$2 cm$^2$ each,
by a system of Cu$-$Be ribbon electrodes, and with no longitudinal
segmentation. The calorimeter is $27X_0$ deep and has an energy
resolution $\sigma(E)/E=0.032/\sqrt{E}\oplus0.09/E\oplus0.0042$ ($E$
in GeV). Spatial resolution for a single electromagnetic shower is
$\sigma_x=\sigma_y=0.42/\sqrt{E}\oplus0.06$ cm for the transverse
coordinates $x$ and $y$.

The LKr is followed by a hadronic calorimeter (HAC) and a muon
detector (MUV), both not used in the present analysis. A detailed
description of the components of the NA48 detector can be found
elsewhere\cite{fa07}. The NA48/2 experiment took data during two
runs in 2003 and 2004, with about 60 days of effective running each.
About $18\times10^9$ events were recorded in total.

In order to simulate the detector response, a detailed
GEANT-based\cite{geant} Monte Carlo (MC) simulation is employed,
which includes full detector geometry and material description,
stray magnetic fields, DCH local inefficiencies and misalignment,
detailed simulation of the kaon beam line, and time variations of
the above throughout the running period. Radiative corrections are
applied to kaon decays using the PHOTOS package\cite{photos}.

\boldmath
\section{$K^\pm\to\pi^\pm e^+e^-$ analysis}
\unboldmath

The $K^\pm\to\pi^\pm e^+e^-$ rate is measured relatively to the
abundant $K^\pm\to\pi^\pm\pi^0_D$ normalization channel (with
$\pi^0_D\to e^+e^-\gamma$). The final states of the signal and
normalization channels contain identical sets of charged particles.
Thus electron and pion identification efficiencies, potentially
representing a significant source of systematic uncertainties,
cancel in the first order.

\subsection{Event selection}

Three-track vertices (compatible with the topology of
$K^\pm\to\pi^\pm e^+e^-$ and $K^\pm\to\pi^\pm\pi^0_D$ decays) are
reconstructed using the Kalman filter algorithm\cite{fr87} by
extrapolation of track segments from the upstream part of the
spectrometer back into the decay volume, taking into account the
measured Earth's magnetic field, stray field due to magnetization of
the vacuum tank, and multiple scattering in the Kevlar window.

A large part of the selection is common to the signal and
normalization modes. It requires a presence of a vertex satisfying
the following criteria.
\begin{itemize}
\item Total charge of the three tracks: $Q=\pm1$.
\item Vertex longitudinal position is inside fiducial decay volume:
$Z_{\rm vertex}>Z_{\rm final~collimator}$.
\item Particle identification is performed using the ratio $E/p$ of
track energy deposition in the LKr to its momentum measured by the
spectrometer. The vertex is required to be composed of one pion
candidate ($E/p<0.85$), and two opposite charge $e^\pm$ candidates
($E/p>0.95$). No discrimination of pions against muons is performed.
\item The vertex tracks are required to be consistent in time
(within a 10~ns time window) and consistent with the trigger time,
to be in DCH, LKr and HOD geometric acceptance, and to have momenta
in the range $5~{\rm GeV}/c<p<50~{\rm GeV}/c$. Track separations are
required to exceed 2~cm in the DCH1 plane to suppress photon
conversions, and to exceed 15~cm in the LKr plane to minimize
particle misidentification due to shower overlaps.
\end{itemize}
If multiple vertices satisfying the above conditions are found, the
one with the best fit quality is considered. The following criteria
are then applied to the reconstructed kinematic variables to select
the $K^\pm\to\pi^\pm e^+e^-$ candidates.
\begin{itemize}
\item $\pi^\pm e^+e^-$ momentum within the beam nominal range:
$54~{\rm GeV}/c<|\vec p_{\pi ee}|<66~{\rm GeV}/c$.
\item $\pi^\pm e^+e^-$ transverse momentum with respect to
the measured beam trajectory: $p_T^2<0.5\times 10^{-3}~({\rm
GeV}/c)^2$.
\item $\pi^\pm e^+e^-$ invariant mass:
$475~{\rm MeV}/c^2<M_{\pi ee}<505~{\rm MeV}/c^2$.
\item Suppression of the $K^\pm\to\pi^\pm\pi^0_D$ background defining
the visible kinematic region: $z=(M_{ee}/M_K)^2>0.08$, which
approximately corresponds to $M_{ee}>140$~MeV/$c^2$.
\end{itemize}
Independently, a presence of a LKr energy deposition cluster (photon
candidate) satisfying the following principal criteria is required
to select the $K^\pm\to\pi^\pm\pi^0_D$ candidates.
\begin{itemize}
\item Cluster energy $E>3$~GeV, cluster time consistent with the
vertex time, sufficient transverse separations from track impact
points at the LKr plane ($R_{\pi\gamma}>30$~cm,
$R_{e\gamma}>10$~cm).
\item $e^+e^-\gamma$ invariant mass compatible with a $\pi^0$
decay: $|M_{ee\gamma}-M_{\pi^0}|<10$~MeV/$c^2$.
\item The same conditions on reconstructed $\pi^\pm e^+e^-\gamma$
total and transverse momenta as used for $\pi^\pm e^+e^-$ momentum
in the $K^\pm\to\pi^\pm e^+e^-$ selection.
\item $\pi^\pm e^+e^-\gamma$ invariant mass:
$475~{\rm MeV}/c^2<M_{\pi ee\gamma}<510~{\rm MeV}/c^2$.
\end{itemize}

\subsection{Signal and normalization samples}

\begin{figure}[t]
  \vspace{57mm}
  \includegraphics{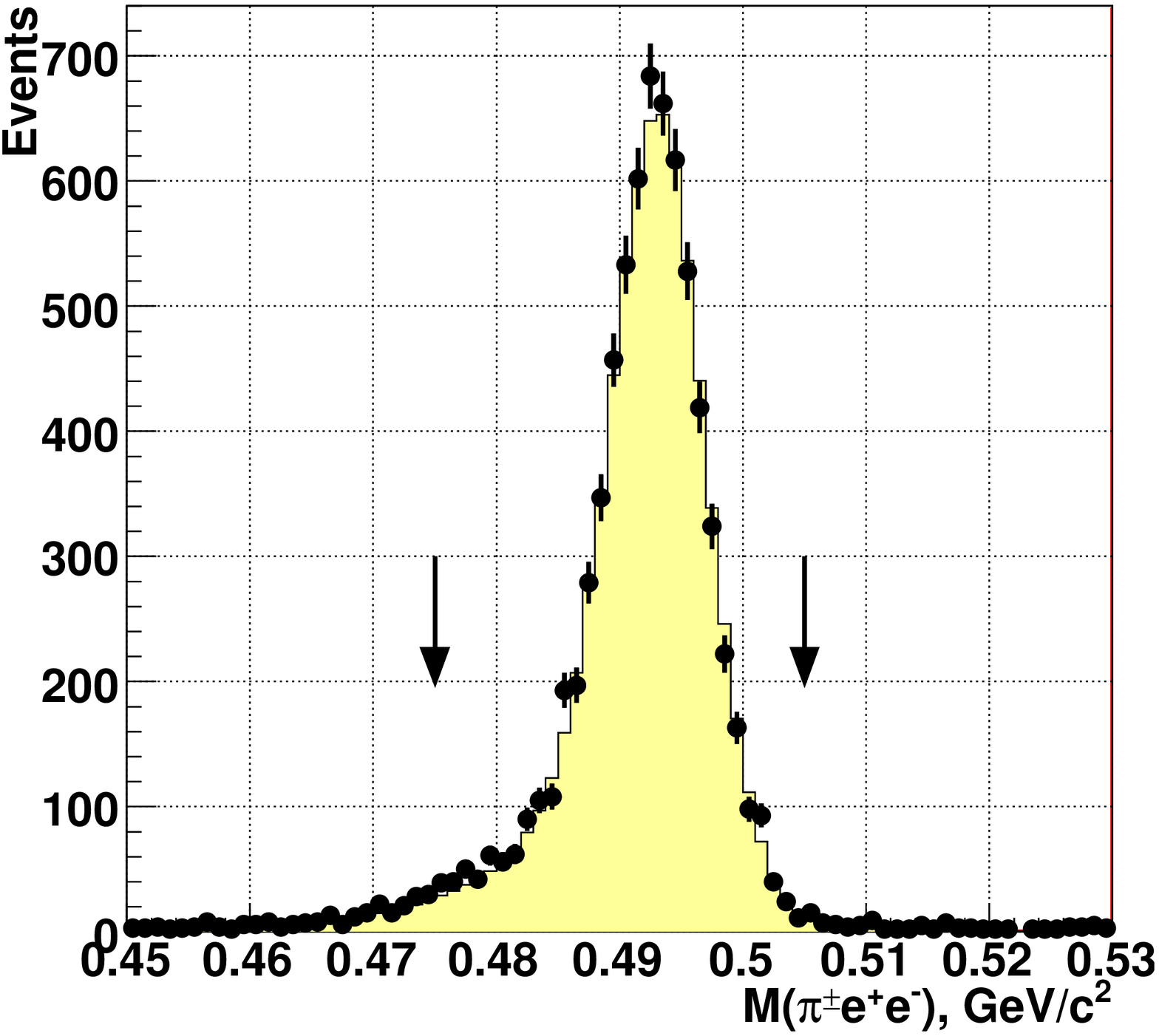}~~~~~~~~~~~~~~~~~~~~~~~~~~~~~~~~~~~~~~~~~~~~~~~~~~
      \includegraphics{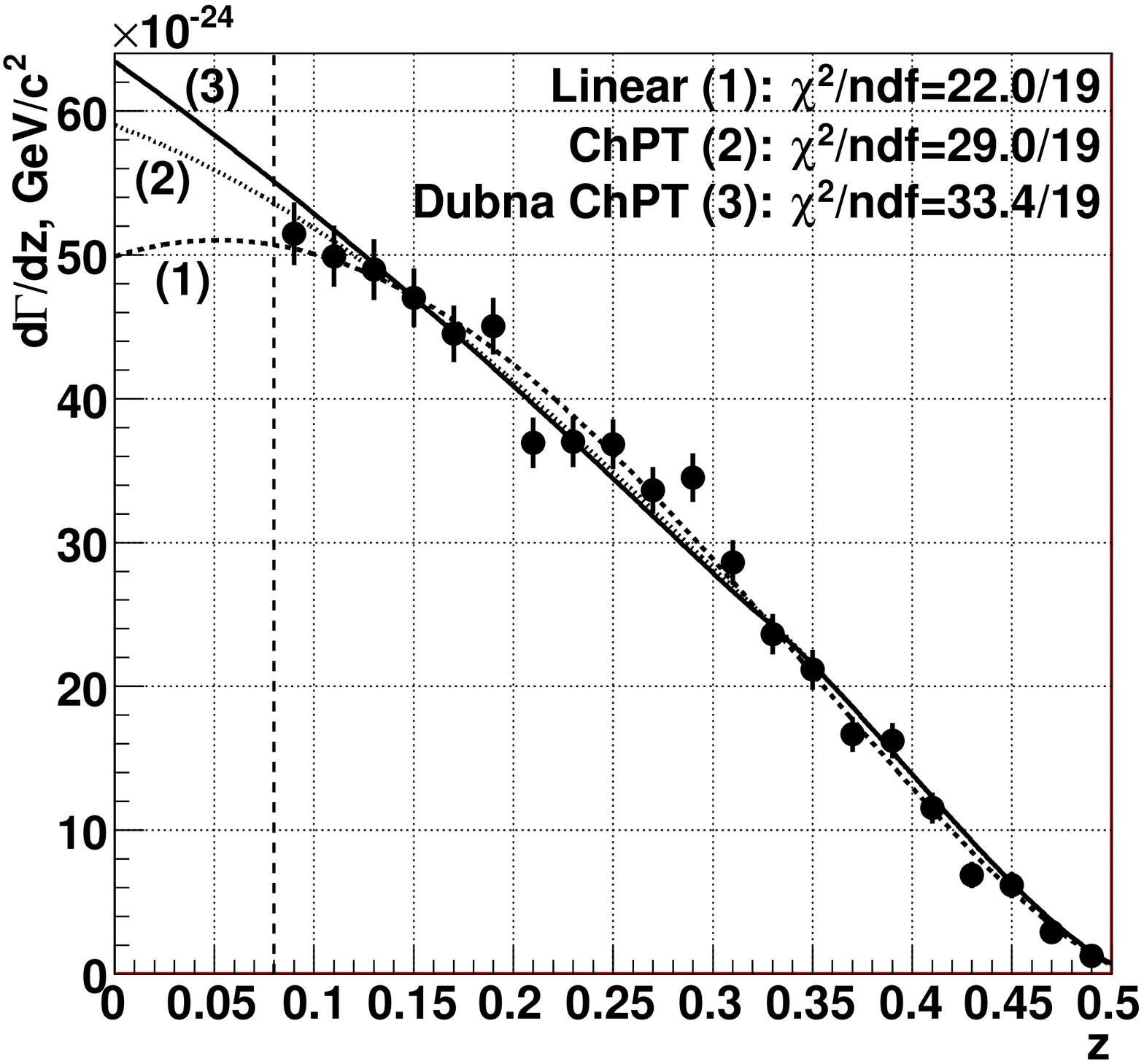}
  \caption{\it
    Left: reconstructed spectrum of $\pi^\pm
e^+e^-$ invariant mass; data (dots) and MC simulation (filled area).
Right: the computed $d\Gamma_{\pi ee}/dz$ (background subtracted,
trigger efficiencies corrected for) and the results of fits
according to the considered models.
    \label{fig:mk}}
\end{figure}

The reconstructed $\pi^\pm e^+e^-$ invariant mass spectrum is
presented in Fig.~\ref{fig:mk} (left plot). The $\pi^\pm e^+e^-$
mass resolution is $\sigma_{\pi ee}=4.2$~MeV/$c^2$, in agreement
with MC simulation. The $e^+e^-$ mass resolution computed by MC
simulation is $\sigma_{ee}=2.3$~MeV/$c^2$.

In total 7,146 $K^\pm\to\pi^\pm e^+e^-$ candidates are found in the
signal region. After the kinematical suppression of the $\pi^0_D$
decays, residual background contamination mostly results from
particle misidentification (i.e. $e^\pm$ identified as $\pi^\pm$ and
vice versa). The following relevant background sources were
identified with MC simulations: (1) $K^\pm\to\pi^\pm\pi^0_D$ with
misidentified $e^\pm$ and $\pi^\pm$; (2) $K^\pm\to\pi^0_De^\pm\nu$
with a misidentified $e^\pm$ from the $\pi^0_D$ decay. Background
estimation by selecting the strongly suppressed\cite{ap00} lepton
number violating $K^\pm\to\pi^\mp e^\pm e^\pm$ (``same-sign'')
candidates was considered the most reliable method. For the above
two background sources, the expected mean numbers and kinematic
distributions of the selected same-sign candidates are identical to
those of background events (up to a negligible acceptance
correction). In total 44 events pass the same-sign selection, which
leads to background estimation of $(0.6\pm0.1)\%$. This result was
independently confirmed with MC simulation of the two background
modes.

In total $12.228\times 10^6$ $K^\pm\to\pi^\pm\pi^0_D$ candidates are
found in the signal region. The only significant background source
is the semileptonic $K^\pm\to\pi^0_D\mu^\pm\nu$ decay. Its
contribution is not suppressed by particle identification cuts,
since no $\pi$/$\mu$ separation is performed. The background
contamination is estimated to be 0.15\% by MC simulation.

\subsection{Trigger chain and its efficiency}

Both $K^\pm\to\pi^\pm e^+e^-$ and $K^\pm\to\pi^\pm\pi^0_D$ samples
(as well as $K^\pm\to3\pi^\pm$) are recorded via the same two-level
trigger chain. At the first level (L1), a coincidence of hits in the
two planes of the HOD in at least two of the 16 non-overlapping
segments is required. The second level (L2) is based on a hardware
system computing coordinates of hits from DCH drift times, and a
farm of asynchronous processors performing fast track reconstruction
and running a selection algorithm, which basically requires at least
two tracks to originate in the decay volume with the closest
distance of approach of less than 5 cm. L1 triggers not satisfying
this condition are examined further and accepted nevertheless if
there is a reconstructed track not kinematically compatible with a
$\pi^\pm\pi^0$ decay of a $K^\pm$ having momentum of 60 GeV/$c$
directed along the beam axis.

The NA48/2 analysis strategy for non-rare decay modes involves
direct measurement of the trigger efficiencies using control data
samples of downscaled low bias triggers collected simultaneously
with the main triggers. However direct measurements are not possible
for the $K^\pm\to\pi^\pm e^+e^-$ events due to very limited sizes of
the corresponding control samples. Dedicated simulations of L1 and
L2 performance (involving, in particular, the measured time
dependencies of local DCH and HOD inefficiencies) were used instead.
The simulated efficiencies and their kinematic dependencies were
compared against measurements for the abundant
$K^\pm\to\pi^\pm\pi^0_D$ and $K^\pm\to\pi^\pm\pi^+\pi^-$ decays in
order to validate the simulations.

The simulated values of L1 and L2 inefficiencies for the selected
$K^\pm\to\pi^\pm\pi^0_D$ sample are $\varepsilon_{L1}=0.37\%$,
$\varepsilon_{L2}=0.80\%$. The values of the integral trigger
inefficiencies for the $K^\pm\to\pi^\pm e^+e^-$ sample depend on the
a priori unknown form factor; the corrections are applied
differentially in bins of dilepton invariant mass. Indicative values
of inefficiencies computed assuming a realistic linear form factor
with a slope $\delta=2.3$ are $\varepsilon_{L1}=0.06\%$,
$\varepsilon_{L2}=0.42\%$. The $K^\pm\to\pi^\pm\pi^0_D$ sample is
affected by larger inefficiencies due to a smaller invariant mass of
the $e^+e^-$ system, which means that the leptons are geometrically
closer.

\subsection{Theoretical input}

The decay is supposed to proceed through one photon exchange,
resulting in a spectrum of the $z=(M_{ee}/M_K)^2$ kinematic variable
sensitive to the form factor $W(z)$\cite{da98}:
\begin{equation}
\frac{d\Gamma}{dz}=\frac{\alpha^2M_K}{12\pi(4\pi)^4}
\lambda^{3/2}(1,z,r_\pi^2)\sqrt{1-4\frac{r_e^2}{z}}
\left(1+2\frac{r_e^2}{z}\right)|W(z)|^2, \label{theory}
\end{equation}
where $r_e=m_e/M_K$, $r_\pi=m_\pi/M_K$, and
$\lambda(a,b,c)=a^2+b^2+c^2-2ab-2ac-2bc$. On the other hand, the
spectrum of the angle $\theta_{\pi e}$ between $\pi$ and $e^+$ in
the $e^+e^-$ rest frame is proportional to $\sin^2\theta_{\pi e}$,
and is not sensitive to $W(z)$.

The following parameterizations of the form factor $W(z)$ are
considered in the present analysis.
\begin{enumerate}
\item Linear: $W(z)=G_FM_K^2f_0(1+\delta z)$
with free normalization and slope $(f_0,\delta)$.
\item Next-to-leading order ChPT\cite{da98}:
$W(z)=G_FM_K^2(a_++b_+z)+W^{\pi\pi}(z)$ with free parameters
$(a_+,b_+)$, and an explicitly calculated pion loop term
$W^{\pi\pi}(z)$.
\item The Dubna version of ChPT parameterization involving meson form
factors: $W(z)\equiv W(M_a,M_\rho,z)$\cite{du06}, with resonance
masses ($M_a$, $M_\rho$) treated as free parameters.
\end{enumerate}
The goal of the analysis is extraction of the form factor parameters
in the framework of each of the above models, and computation of the
corresponding branching fractions ${\rm BR}_{1,2,3}$.

\subsection{Fitting procedure}

The values of $d\Gamma_{\pi ee}/dz$ in the centre of each $i$-bin of
$z$, which can be directly confronted to the theoretical predictions
(\ref{theory}), are then computed as
\begin{equation}
(d\Gamma_{\pi ee}/dz)_i = \frac{N_i-N^B_i}{N_{2\pi}}\cdot
\frac{A_{2\pi}(1-\varepsilon_{2\pi})}{A_i(1-\varepsilon_i)} \cdot
{\rm BR}(K^\pm\to\pi^\pm\pi^0)\cdot{\rm BR}(\pi^0_D)\cdot
\frac{\Gamma_K}{\Delta z}.
\end{equation}
Here $N_i$ and $N^B_i$ are the numbers of observed $K^\pm\to\pi^\pm
e^+e^-$ candidates and background events in the $i$-th bin,
$N_{2\pi}$ is the number of $K^\pm\to\pi^\pm\pi^0_D$ events
(background subtracted), $A_i$ and $\varepsilon_i$ are geometrical
acceptance and trigger inefficiency in the $i$-th bin for the signal
sample (computed by MC simulation), $A_{2\pi}=2.94\%$ and
$\varepsilon_{2\pi}=1.17\%$ are those for $K^\pm\to\pi^\pm\pi^0_D$
events, $\Gamma_K$ is the nominal kaon width\cite{pdg}, $\Delta z$
is the chosen width of the $z$ bin, ${\rm
BR}(K^\pm\to\pi^\pm\pi^0)=(20.64\pm0.08)\%$ (FlaviaNet
average\cite{an08}), ${\rm BR}(\pi^0_D)=(1.198\pm0.032)\%$ (PDG
average\cite{pdg}).

The computed values of $d\Gamma_{\pi ee}/dz$ vs $z$ are presented in
Fig.~\ref{fig:mk} (right plot) along with the results of the fits to
the three considered models. ${\rm BR}(K^\pm\to\pi^\pm e^+e^-)$ in
the full kinematic range corresponding to each model are then
computed using the measured parameters, their statistical
uncertainties, and correlation matrices.

In addition, a model-independent branching fraction ${\rm BR_{mi}}$
in the visible kinematic region $z>0.08$ is computed by integration
of $d\Gamma_{\pi ee}/dz$. ${\rm BR_{mi}}$ is to a good approximation
equal to each of the model-dependent BRs computed in the restricted
kinematic range $z>0.08$.

\subsection{Systematic uncertainties}

The following sources of systematic uncertainties were studied.

1. Particle identification. Imperfect MC description of electron and
pion identification inefficiencies $f_e$ and $f_\pi$ can bias the
result only due to the momentum dependence of the inefficiencies,
due to identical charged particle composition, but differing
momentum distributions of the signal and normalization final states.
Inefficiencies were measured for the data to vary depending on
particle momentum in the ranges $1.6\%<f_\pi<1.7\%$ and
$1.1\%<f_e<1.7\%$ in the analysis track momentum range. Systematic
uncertainties due to these momentum dependencies not perfectly
described by MC were conservatively estimated assuming that MC
predicts momentum-independent $f_e$ and $f_\pi$.

2. Beam line description. Despite the careful simulation of the
beamline including time variations of its parameters, the residual
discrepancies of data and MC beam geometries and spectra bias the
results. To evaluate the related systematic uncertainties,
variations of the results with respect to variations of cuts on
track momenta, LKr cluster energies, total and transverse momenta of
the final states $\pi^\pm e^+e^-(\gamma)$, and track distances from
beam axis in DCH planes were studied.

3. Background subtraction. As discussed above, the same-sign event
spectrum is used for background estimation in the $\pi^\pm e^+e^-$
sample. The method has a limited statistical precision (with an
average of 2 same-sign event in a bin of $z$). Furthermore, the
presence of the component with two $e^+e^-$ pairs (due to both
$\pi^0_D$ decays and external conversions) with a non-unity expected
ratio of same-sign to background events biases the method. The
uncertainties of the measured parameters due to background
subtraction were conservatively taken to be equal to the corrections
themselves.

4. Trigger efficiency. As discussed earlier, the corrections for
trigger inefficiencies were evaluated by simulations. In terms of
decay rates, L1 and L2 corrections have similar integral magnitudes
of a few $10^{-3}$. No uncertainty was ascribed to the L1
correction, due to relative simplicity of the trigger condition. On
the other hand, the uncertainty of the L2 efficiency correction was
conservatively taken to be equal to the correction itself.

5. Radiative corrections. Uncertainties due to the radiative
corrections were evaluated by variation of the lower $\pi^\pm
e^+e^-$ invariant mass cut.

6. Fitting method. Uncertainties due to the fitting procedure were
evaluated by variation of the $z$ bin width.

7. External input. Substantial uncertainties arise from the external
input, as ${\rm BR}(\pi^\pm\pi^0_D)$ is experimentally known only
with 2.7\% relative precision\cite{pdg}. The only parameter not
affected by an external uncertainty is the linear form factor slope
$\delta$ describing only the shape of the spectrum.

\begin{table}[tb]
\begin{center}
\begin{tabular}{@{}r@{~~}r@{~~}r@{~~}r@{~$\pm$~}l@{~}r@{~$\pm$~}l@{~~}r@{~~}r}
\hline Parameter&$e,\pi$&Beam      &\multicolumn{2}{c}{Background} &\multicolumn{2}{c}{Trigger}   &Rad. &Fitting\\
                &ID     &spectra   &\multicolumn{2}{c}{subtraction}&\multicolumn{2}{c}{efficiency}&corr.&method\\
                \hline
$\delta$ & 0.01& 0.04&$-0.04$ &0.04 &$-0.03$ &0.03  & 0.05&0.03\\
$f_0$    &0.001&0.006&$0.002$ &0.002&$0.000$ &0.001 &0.006&0.003\\
$a_+$    &0.001&0.005&$-0.001$&0.001&$-0.001$&0.002 &0.005&0.004\\
$b_+$    &0.009&0.015&$0.017$ &0.017&$0.016$ &0.015 &0.015&0.010\\
$M_a$/GeV&0.004&0.009&$0.008$ &0.008&$0.006$ &0.006 &0.009&0.006\\
$M_b$/GeV&0.002&0.003&$0.003$ &0.003&$0.003$ &0.003 &0.004&0.002\\
\hline
${\rm BR}_{1,2,3}\!\!\times\!\!10^7$&0.02&0.02&$-0.01$&0.01&$-0.02$&0.01&0.01&0.02\\
${\rm BR_{mi}}\!\!\times\!\!10^7$   &0.02&0.01&$-0.01$&0.01&$-0.02$&0.01&0.01&n/a\\
\hline
\end{tabular}
\end{center}
\vspace{-5mm} \caption{Summary of corrections and systematic
uncertainties (excluding the external ones).} \label{tab:syst}
\end{table}

The applied corrections and the systematic uncertainties (excluding
the external ones presented later) are summarized in
Table~\ref{tab:syst}.

\subsection{Results and discussion}

The measured parameters of the considered models and the
corresponding BRs in the full $z$ range, as well the
model-independent ${\rm BR_{mi}}(z>0.08)$, with their statistical,
systematic, and external uncertainties are presented in
Table~\ref{tab:results}. The correlation coefficients between the
pairs of model parameters, not listed in the table, are
$\rho(\delta,f_0)=-0.963$, $\rho(a_+,b_+)=-0.913$, and
$\rho(M_a,M_\rho)=0.998$.

\begin{table}[tb]
\begin{center}
\begin{tabular}{@{}r@{~$=$~}r@{~$\pm$~}r@{~$\pm$~}r@{~$\pm$~}r@{~$=$~}r@{~$\pm$~}r@{}}
\hline
$\delta$                        &$2.35$  &$0.15_{\rm stat.}$ &$0.09_{\rm syst.}$ &$0.00_{\rm ext.}$ &$2.35$ &0.18\\
$f_0$                           &$0.532$ &$0.012_{\rm stat.}$&$0.008_{\rm syst.}$&$0.007_{\rm ext.}$&$0.532$&0.016\\
${\rm BR}_1\times10^7$          &$3.02$  &$0.04_{\rm stat.}$ &$0.04_{\rm syst.}$ &$0.08_{\rm ext.}$ &$3.02$ &0.10\\
\hline
$a_+$                           &$-0.579$&$0.012_{\rm stat.}$&$0.008_{\rm syst.}$&$0.007_{\rm ext.}$&$-0.579$&0.016\\
$b_+$                           &$-0.798$&$0.053_{\rm stat.}$&$0.037_{\rm syst.}$&$0.017_{\rm ext.}$&$-0.798$&0.067\\
${\rm BR}_2\times10^7$          &$3.11$  &$0.04_{\rm stat.}$ &$0.04_{\rm syst.}$ &$0.08_{\rm ext.}$ &$3.11$  &0.10\\
\hline
$M_a/{\rm GeV}$                 &$0.965$ &$0.028_{\rm stat.}$&$0.018_{\rm syst.}$&$0.002_{\rm ext.}$&$0.965$&0.033\\
$M_\rho/{\rm GeV}$              &$0.711$ &$0.010_{\rm stat.}$&$0.007_{\rm syst.}$&$0.002_{\rm ext.}$&$0.711$&0.013\\
${\rm BR}_3\times10^7$          &$3.15$  &$0.04_{\rm stat.}$ &$0.04_{\rm syst.}$ &$0.08_{\rm ext.}$ &$3.15$ &0.10\\
\hline
${\rm BR_{mi}}\times10^7$&$2.26$  &$0.03_{\rm stat.}$ &$0.03_{\rm syst.}$ &$0.06_{\rm ext.}$ &$2.26$ &0.08\\
\hline
\end{tabular}
\end{center}
\vspace{-5mm} \caption{Results of fits to the three considered
models, and the model-independent ${\rm BR_{mi}}(z>0.08)$.}
\label{tab:results}
\end{table}

Fits to all the three models are of reasonable quality, however the
linear form-factor model leads to the smallest $\chi^2$. The data
sample is insufficient to distinguish between the models considered.

The obtained form factor slope $\delta$ is in agreement with the
previous measurements based on $K^+\to\pi^+e^+e^-$\cite{al92,ap99}
and $K^\pm\to\pi^\pm\mu^+\mu^-$\cite{ma00} samples, and further
confirms the contradiction of the data to meson dominance
models\cite{li99}. The obtained $f_0$, $a_+$ and $b_+$ are in
agreement with the only previous measurement\cite{ap99}. The
measured parameters $M_a$ and $M_\rho$ are a few \% away from the
nominal masses of the resonances\cite{pdg}.

The branching ratio in the full kinematic range, which is computed
as the average between the two extremes corresponding to the models
(1) and (3), and includes an uncertainty due to extrapolation into
the inaccessible region $z<0.08$, is
\begin{displaymath}
{\rm BR}\!=\!(3.08\pm0.04_{\rm stat.}\pm0.04_{\rm syst.}\pm0.08_{\rm
ext.}\pm0.07_{\rm model})\times10^{-7}\!=\!
(3.08\pm0.12)\times10^{-7}.
\end{displaymath}
It should be stressed that a large fraction of the uncertainty of
this result is correlated with the earlier measurements. A
comparison to the precise BNL E865 measurement\cite{ap99} dismissing
correlated uncertainties due to external BRs and model dependence,
and using the same external input, shows a $1.4\sigma$ difference.
In conclusion, the obtained BR is in agreement with the previous
measurements.

Finally, a first measurement of the direct CP violating asymmetry of
$K^+$ and $K^-$ decay rates in the full kinematic range was obtained
by performing BR measurements separately for $K^+$ and $K^-$ and
neglecting the correlated uncertainties: $\Delta(K_{\pi
ee}^\pm)=({\rm BR}^+-{\rm BR}^-)/({\rm BR}^++{\rm
BR}^-)=(-2.1\pm1.5_{\rm stat.}\pm 0.3_{\rm syst.})\%$. The result is
compatible to no CP violation. However its precision is far from the
theoretical expectation\cite{da98} of $|\Delta(K_{\pi ee}^\pm)|\sim
10^{-5}$.

\boldmath
\section{$K^\pm\to\pi^\pm\gamma\gamma$ analysis}
\unboldmath

The $K^\pm\to\pi^\pm\gamma\gamma$ rate is measured relatively to the
$K^\pm\to\pi^\pm\pi^0$ normalization channel. The signal and
normalization channels have identical particle composition of the
final states, and the only cut differing for the two channels is the
one on the $\gamma\gamma$ invariant mass. The used trigger chain
involves the so called ``neutral trigger'' based on requirement of
minimal number of energy deposition clusters in the LKr calorimeter.

\begin{figure}[t]
  \vspace{60mm}
  \includegraphics{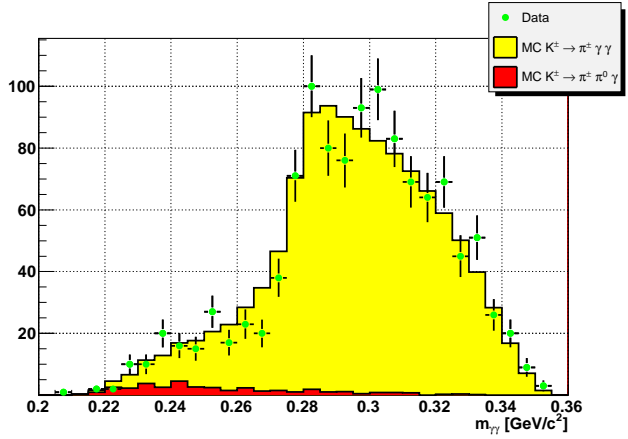}
  \caption{\it
    The reconstructed spectrum of $\gamma\gamma$ invariant mass for
    the $K^\pm\to\pi^\pm\gamma\gamma$ decay
    (dots), and its comparison to MC expectation assuming
    ChPT ${\cal O}(p^6)$ distribution with $\hat c=2$ (filled area).
    \label{fig:pigg}}
\end{figure}

About 40\% of the total NA48/2 data sample have been analyzed, and
1,164 $K^\pm\to\pi^\pm\gamma\gamma$ decay candidates (with
background contamination estimated by MC to be 3.3\%) are found,
which has to be compared with the only previous
measurement\cite{ki97} involving 31 decay candidates. The
reconstructed spectrum of $\gamma\gamma$ invariant mass in the
accessible kinematic region $M_{\gamma\gamma}>0.2$~GeV/c$^2$ is
presented in Fig.~\ref{fig:pigg}, along with a MC expectation
assuming ChPT ${\cal O}(p^6)$ distribution\cite{da96} with a
realistic parameter $\hat c=2$. ChPT predicts an enhancement of the
decay rate (cusp-like behaviour) at the $\pi\pi$ mass threshold
$m_{\gamma\gamma}\approx280$~MeV/c$^2$, independently of the value
of the $\hat c$ parameter. The observed spectrum provides the first
clean experimental evidence for this phenomenon.

As the first step of the analysis, the partial width of the decay
was measured assuming the ChPT ${\cal O}(p^6)$ shape with a fixed
parameter $\hat c=2$. The following preliminary result, which is in
agreement with the ChPT computation for $\hat c=2$, was obtained:
\begin{displaymath}
{\rm BR}=(1.07\pm0.04_{\rm stat.}\pm0.08_{\rm syst.})\times 10^{-6}.
\end{displaymath}
A combined fit of the $m_{\gamma\gamma}$ spectrum shape and the
decay rate is foreseen to measure the $\hat c$ parameter.

\boldmath
\section{$K^\pm\to\pi^\pm\gamma e^+e^-$ analysis}
\unboldmath

The $K^\pm\to\pi^\pm\gamma e^+e^-$ rate is measured relatively to
the $K^\pm\to\pi^\pm\pi^0_D$ normalization channel. The signal and
normalization channels have identical particle composition of the
final states. The same trigger chain as for the collection of
$K^\pm\to\pi^\pm e^+e^-$ is used.

\begin{figure}[t]
  \vspace{55mm}
      \includegraphics{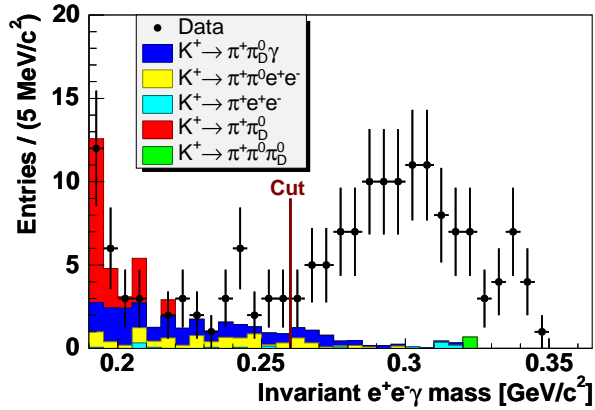}
  \caption{\it
    The reconstructed spectrum of $\gamma e^+e^-$ invariant mass for
    the $K^\pm\to\pi^\pm\gamma e^+e^-$ decay
    (dots), and MC background expectations (filled areas).
    \label{fig:pigee}}
\end{figure}

With the full NA48/2 data sample analyzed, 120
$K^\pm\to\pi^\pm\gamma e^+e^-$ decay candidates (with the background
estimated by MC to be 6.1\%) are found in the accessible kinematic
region $M_{\gamma ee}>0.26$~GeV/c$^2$. This is the first observation
of this decay mode. The reconstructed spectrum of $\gamma e^+e^-$
invariant mass is presented in Fig.~\ref{fig:pigee}, along with MC
expectations for background contributions. The spectrum provides
another evidence for the rate enhancement at the $\pi\pi$ mass
threshold.

The final results of the analysis have recently been
published\cite{ba08}. The model-independent partial width in the
accessible kinematic region is measured to be
\begin{displaymath}
{\rm BR}(M_{\gamma ee}>0.26~{\rm GeV}/c^2)=(1.19\pm0.12_{\rm
stat.}\pm0.04_{\rm syst.})\times 10^{-8}.
\end{displaymath}
The ChPT parameter $\hat c$ assuming ${\cal O}(p^4)$
distibution\cite{ga99} was measured to be $\hat c=0.90\pm0.45$.

\section*{Conclusions}

A precise study of the $K^\pm\to\pi^\pm e^+e^-$ decay has been
performed. The data sample and precision are comparable to world's
best ones, the preliminary results are in agreement with the
previous measurements, and the first limit on CP violating charge
asymmetry has been obtained.

A precise study of the $K^\pm\to\pi^\pm\gamma\gamma$ has been
performed. The first clear evidence for a rate enhancement at
$\pi\pi$ mass threshold has been obtained. The preliminary
measurement of BR agrees with the ChPT prediction. A detailed
spectrum shape study is foreseen.

The first observation of the $K^\pm\to\pi^\pm\gamma e^+e^-$ decay,
and measurement of its parameters, including the BR, have been
performed. The $M_{\gamma ee}$ spectrum provides an independent
evidence for the cusp at the $\pi\pi$ mass threshold.


\begin{thebibliography}{99}
%
\bibitem{ek87}
G. Ecker, A. Pich, E. de Rafael, Nucl. Phys. {\bf B291} (1987) 692.
%
\bibitem{da98}
G. D'Ambrosio {\it et al.}, JHEP {\bf 8} (1998) 4.
%
\bibitem{du06}
A.Z. Dubni\v ckov\'a {\it et al.}, Phys. Part. Nucl. Lett. {\bf 5},
vol. 2 (2008) 76 [hep-ph/0611175].
%
\bibitem{bl75}
P. Bloch {\it et al.}, Phys. Lett. {\bf B56} (1975) 201 (1975).
%
\bibitem{al92}
C. Alliegro {\it et al.}, Phys. Rev. Lett. {\bf 68} (1992) 278
(1992).
%
\bibitem{ap99}
R. Appel {\it et al.}, Phys. Rev. Lett. {\bf 83} (1999) 4482.
%
\bibitem{da96}
G. D'Ambrosio, J. Portol\'es, Phys. Lett. {\bf B386} (1996) 403.
%
\bibitem{ga99}
F. Gabbiani, Phys. Rev. {\bf D59} (1999) 094022.
%
\bibitem{ki97}
P. Kitching {\it et al.}, Phys. Rev. Lett. {\bf 79} (1997) 4079.
%
\bibitem{ba08}
J.R. Batley {\it et al.}, Phys. Lett. {\bf B659} (2008) 493.
%
\bibitem{ba07}
J.R. Batley {\it et al.}, Eur. Phys. J. {\bf C52} (2007) 875.
%
\bibitem{ba96}
G.D. Barr {\it et al.}, \NIMA{370}{1996}{413}.
%
\bibitem{fa07}
V. Fanti {\it et al.}, \NIMA{574}{2007}{433}.
%
\bibitem{geant}
GEANT detector description and simulatio tool, CERN program library
long writeup W5013 (1994).
%
\bibitem{photos}
E. Barberio and Z. Was, Comp. Phys. Comm. {\bf 79} (1994) 291.
%
\bibitem{fr87}
R. Fr\"uhwirth, \NIMA{262}{1987}{444}.
%
\bibitem{ap00}
R. Appel {\it et al.}, Phys. Rev. Lett. {\bf 85} (2000) 2877.
%
\bibitem{an08}
M. Antonelli {\it et al.}, arXiv:0801.1817.
%
\bibitem{pdg}
W.-M. Yao {\it et al.} (PDG), J. Phys. {\bf G33} (2006) 1.
%
\bibitem{ma00}
H. Ma {\it et al.}, Phys. Rev. Lett. {\bf 84} (2000) 2580.
%
\bibitem{li99}
P. Lichard, Phys. Rev. {\bf D60} (1999) 053007.
\end{thebibliography}
\end{document}